\documentclass[floatfix, aps, amsmath, nofootinbib, twocolumn, 10pt]{revtex4}
\usepackage{xcolor}
\usepackage{listings}
\usepackage{graphicx}
\usepackage{bm}
\usepackage{rotating}
\usepackage{array}
\usepackage{amsmath}
\usepackage{amssymb}
\usepackage{mathrsfs}
\usepackage{cancel}
\usepackage{subfig}
\usepackage{float}
\usepackage{caption}
\graphicspath{{./}}

\def\({\left(}
\def\){\right)}
\def\[{\left[}
\def\]{\right]}

\def\e{\begin{equation}}
\def\q{\end{equation}}
\def\m{\begin{eqnarray}}
\def\n{\end{eqnarray}}

\begin{document}
\title{A Neutron Star Hidden Inside a Black Hole}
\author{Chen Tan$^{1, 2}$}
\author{Yong-Qiang Wang$^{1, 2}$}
\thanks{Corresponding author: {yqwang@lzu.edu.cn}}
\affiliation{$^{1}$Lanzhou Center for Theoretical Physics, Key Laboratory of Theoretical Physics of Gansu Province,
	School of Physical Science and Technology, Lanzhou University, Lanzhou 730000, China}
\affiliation{$^{2}$Institute of Theoretical Physics $\&$ Research Center of Gravitation, Lanzhou University, Lanzhou 730000, China}

\date{\today}

\begin{abstract}
We investigate dark matter admixed neutron stars in which the neutron star coexists with an anisotropic dark matter halo described by the Einasto density profile, a model recently shown to produce regular, singularity-free black hole solutions~[Phys. Rev. D \textbf{113}, 043011 (2026)].
Solving the modified Tolman--Oppenheimer--Volkoff equations with two different equations of state (BSk19 and SLy4), we find that the dark matter halo significantly alters the neutron star structure.
Furthermore, for a specific range of halo parameters, $g_{rr}^{-1}$ changes sign outside the stellar surface, forming an event horizon with the neutron star persisting as a regular configuration inside it--- ``neutron stars in black holes''.
This configuration appears in both equations of state and does not depend on a specific choice of the equation of state.
The discovery of this configuration provides a new perspective and a concrete computable instance for the study of what lies inside a black hole.
\end{abstract}

\maketitle

\section{Introduction}
\label{sec:intro}

The nature of dark matter remains a major open question in modern physics. Its gravitational evidence is inferred from the flat rotation curves of spiral galaxies~\cite{Rubin:1970zza,Rubin:1978kmz,Sofue:2000jx}, the dissociation of luminous and gravitational mass in merging clusters such as the Bullet Cluster~\cite{Clowe:2006eq}, and the pattern of temperature anisotropies in the cosmic microwave background~\cite{Planck:2018vyg}, which collectively indicate that approximately 85\% of the matter in the Universe is non-luminous and non-baryonic~\cite{Bertone:2004pz,Bertone:2016nfn}.
Yet despite this wealth of gravitational evidence, the particle identity of dark matter remains unknown.
Direct detection experiments~\cite{XENON:2018voc,PandaX-4T:2021bab}, collider searches at the LHC~\cite{ATLAS:2021kxv}, and indirect astrophysical probes~\cite{Fermi-LAT:2015att} have steadily tightened exclusion limits without a conclusive signal~\cite{Schumann:2019eaa}.
This impasse has also motivated complementary strategies that exploit the unique physical conditions of extreme astrophysical environments to search for and study dark matter.

Neutron stars are a prominent example of such an environment~\cite{Shao:2022koz}, offering both extreme physical conditions and a rich set of observational channels.
Formed in the aftermath of core-collapse supernovae, they compress roughly 1.4 to 2.5 solar masses of baryonic matter into a sphere of radius approximately 10--12~km~\cite{Lattimer:2004pg,Lattimer:2006xb,Antoniadis:2013pzd,Oertel:2016bki}, producing central densities that exceed nuclear saturation by factors of several~\cite{Lattimer:2015nhk}.
Their compactness $GM/(R c^2)$ reaches $0.1$--$0.3$, placing them firmly in the strong-field regime of general relativity.
This combination of supranuclear density and strong gravity, together with their long lifetimes of up to $10^{10}$ years, allows neutron stars to accumulate dark matter over cosmic timescales~\cite{Bertone:2007ae,Kouvaris:2010vv,Ellis:2018bkr}.
The advent of GW170817~\cite{LIGOScientific:2017zic} opened the era of multi-messenger neutron star astrophysics, complemented by precise mass--radius measurements from the NICER mission~\cite{Miller:2025qfq,Riley:2019yda}. These observational advances provide a foundation for identifying dark matter effects in neutron stars~\cite{Grippa:2024ach}.

Dark matter admixed neutron stars (DANSs) have been extensively
studied~\cite{Grippa:2024ach}. The standard framework is the two-fluid
Tolman--Oppenheimer--Volkoff (TOV) formalism~\cite{Tolman:1939jz,Oppenheimer:1939ne},
which has been applied primarily to two classes of dark matter
candidates.
Bosonic dark matter has been widely
studied~\cite{Giangrandi:2022wht,
Karkevandi:2021ygv,Karkevandi:2024vov,Leung:2022wcf,Jockel:2023rrm},
yielding configurations from compact cores to extended halos.
For fermionic dark matter, models range from a free Fermi gas to
interacting scenarios with scalar and vector
mediators~\cite{Xiang:2013xwa,Ellis:2018bkr,Das:2020ecp,Grippa:2024sfu,Arvikar:2025dwl,Cano:2026yrl}.
Mirror dark matter~\cite{Sandin:2008db} has also been investigated
within this framework.
Subsequent studies have further extended these models to ultracompact
configurations, quasinormal modes, finite-temperature effects,
asteroseismology, stability theory, gravitational synchronization,
and observational signatures~\cite{Pitz:2024xvh,
Boumaza:2026ewo,Mukherjee:2025omu,Sotani:2025lzy,
Zhou:2025kku,Lazarte:2025etl,Sun:2023cqr,
Hu:2025eyt,Zhou:2025dmy}.

Apart from the dark matter types discussed above, a distinct dark matter configuration has recently been explored by
Konoplya and Zhidenko~\cite{Konoplya:2025ect}. This theoretical dark matter possesses unique properties that allow it to serve as the matter source for regular black holes.
Adopting the relation $p_{r{\rm (d)}} = -\rho_{\rm d} \,c^2$ between radial pressure and
density, and taking the Einasto~\cite{Einasto:1965czb} and
Dehnen-type~\cite{Dehnen:1993} dark matter profiles, they found
asymptotically flat, singularity-free black hole solutions, i.e., regular black holes.

\begin{figure}[t]
\centering
\includegraphics[width=\columnwidth]{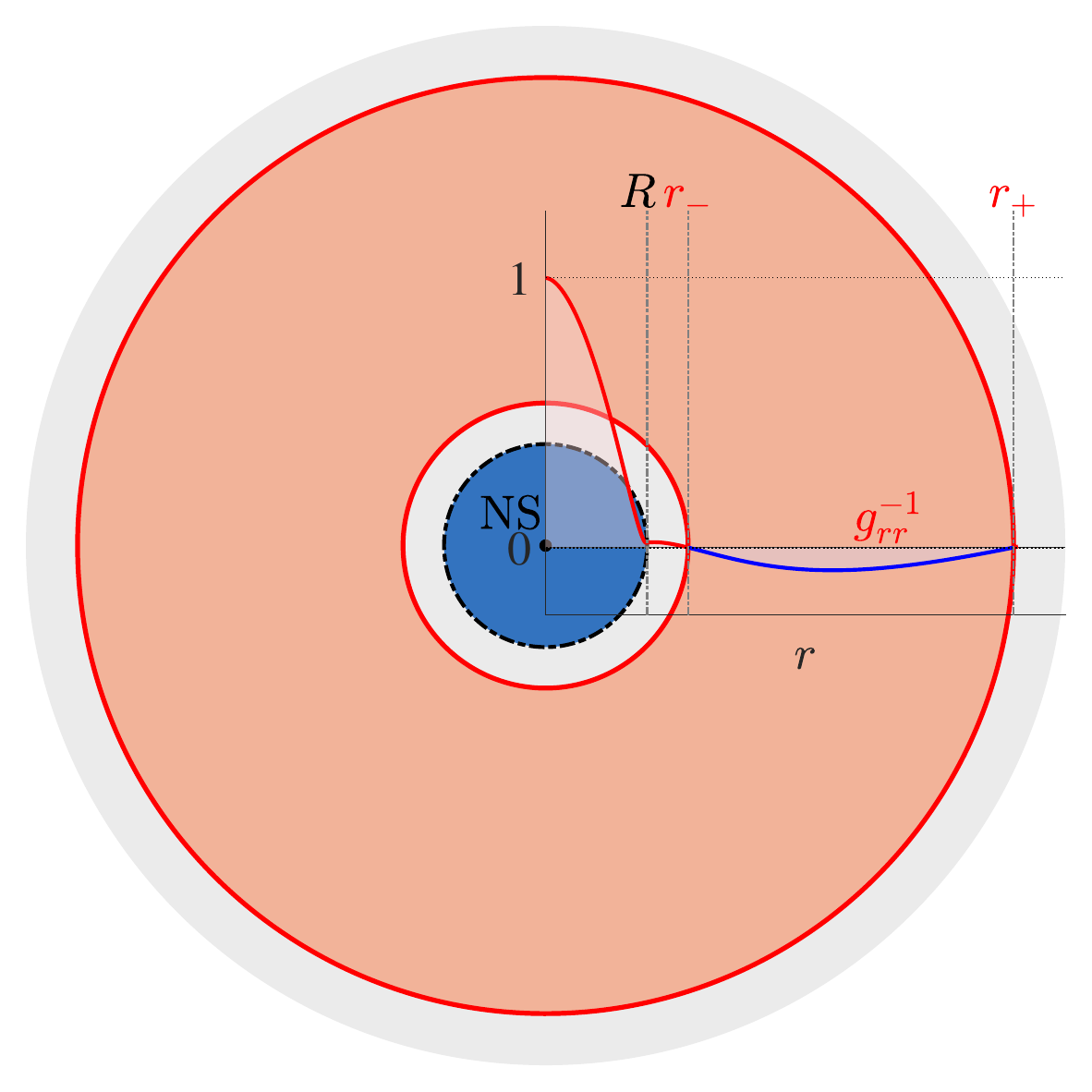}
\caption{Schematic illustration of a neutron star in a black hole. The neutron star (blue) resides in the regular interior, surrounded by a region where $g_{rr}^{-1}<0$ (orange), bounded by an inner ($r_-$) and outer ($r_+$) horizon. The inset shows $g_{rr}^{-1}$ as a function of the radial coordinate.}
\label{fig:schematic}
\end{figure}

In this paper, we investigate a neutron star coexisting with such a dark matter halo. We find that the dark matter halo significantly alters the neutron star structure, and that for a specific range of halo parameters, $g_{rr}^{-1}$ changes sign outside the stellar surface: it remains positive throughout the neutron star interior, drops below zero in a radial shell, and returns to positive values at larger distances. This sign change signals the formation of an event horizon, with the neutron star persisting as a regular, non-singular configuration fully contained within it, a ``neutron star in a black hole,'' as illustrated schematically in Fig.~\ref{fig:schematic}. This phenomenon occupies a specific parameter window: if the halo is too diffuse, no horizon forms; if it is too dense, hydrostatic equilibrium breaks down.

The remainder of this paper is organized as follows.
Sec.~\ref{sec:framework} develops the theoretical framework: the two-component action, the dark matter model with the Einasto profile and $p_{r{\rm (d)}} = -\rho_{\rm d} \,c^2$ equation of state, and the derivation of the modified Tolman--Oppenheimer--Volkoff equations.
Sec.~\ref{sec:numerical} presents our numerical results: we first examine the non-monotonic influence of the three Einasto parameters on the structure of dark matter admixed neutron stars, then demonstrate the gradual formation of the configuration and map its phase diagram, compute the mass--radius relations for two equations of state, and verify the numerical robustness of the solution.
We conclude in Sec.~\ref{sec:discussion} with a summary of our findings and a discussion of their implications.

\section{Framework}
\label{sec:framework}

\subsection{The Model}

We consider a static, spherically symmetric spacetime containing both ordinary baryonic matter and a dark matter component. We assume no interaction between the two sectors except through gravity. This assumption is justified by the latest constraints from dark matter direct detection experiments and the Bullet Cluster~\cite{Clowe:2006eq,Randall:2008ppe}, which show that the dark matter--baryon cross section is many orders of magnitude below the typical nuclear one, $\sigma_\chi \sim 10^{-45}\,\text{cm}^2 \ll \sigma_N \sim 10^{-24}\,\text{cm}^2$. Consequently, the stress-energy tensors of both components are conserved independently, $\nabla_\mu T^{\mu\nu}_{\rm (b)} = 0$ and $\nabla_\mu T^{\mu\nu}_{\rm (d)} = 0$.

The total action is
\begin{equation}
\label{act}
S = \int \sqrt{-g} \, d^4 x \left( \frac{c^3}{16\pi G} R + \frac{1}{c} \mathcal{L}_{\rm b} + \frac{1}{c} \mathcal{L}_{\rm d} \right),
\end{equation}
where $R$ denotes the scalar curvature, $\mathcal{L}_{\rm b}$ is the Lagrangian of baryonic matter, and $\mathcal{L}_{\rm d}$ is the Lagrangian of the dark matter.

Varying the action with respect to the metric yields the Einstein equations:
\begin{equation}
\label{Gx}
R_{\mu\nu} - \frac{1}{2} g_{\mu\nu} R = \frac{8\pi G}{c^4} \left( T_{\mu\nu}^{\rm (b)} + T_{\mu\nu}^{\rm (d)} \right).
\end{equation}

For the dark matter, following Ref.~\cite{Konoplya:2025ect}, we adopt an anisotropic energy-momentum tensor:
\begin{align}
T^{t}_{t{\rm (d)}} &= -c^2 \rho_{\rm d}(r), \label{Ttt} \\
T^{r}_{r{\rm (d)}} &= p_{r{\rm (d)}}(r), \label{Trr} \\
T^{\theta}_{\theta{\rm (d)}} = T^{\varphi}_{\varphi{\rm (d)}} &= p_{\perp{\rm (d)}}(r). \label{Ttheta}
\end{align}

The dark matter density profile is taken to be the Einasto distribution~\cite{Einasto:1965czb}:
\begin{equation}
\label{einasto}
\rho_{\rm d}(r) = \rho_0 \, \exp\!\left[-\left(\frac{r}{h}\right)^{1/n}\right], \quad n > 0,
\end{equation}
where $\rho_0$ is the central dark matter density, $h$ is the characteristic scale radius, and $n$ is the Einasto index. 
The regularity of the central region is ensured by the equation of state for the radial pressure~\cite{Konoplya:2025ect}:
\begin{equation}
\label{pr}
p_{r{\rm (d)}}(r) = -\rho_{\rm d}(r) \, c^2.
\end{equation}
The transverse pressure $p_{\perp{\rm (d)}}$ is then determined by the conservation equation $\nabla_\mu T^{\mu\nu}_{\rm (d)} = 0$, which yields
\begin{equation}
\label{pv}
p_{\perp{\rm (d)}}(r) = -\frac{c^2}{2} \left[ r \rho_{\rm d}'(r) + 2 \rho_{\rm d}(r) \right].
\end{equation}
With $p_{r{\rm (d)}} = -\rho_{\rm d} c^2$ and the above expression for $p_{\perp{\rm (d)}}$, the weak energy condition $\rho_{\rm d} c^2 + p_i \geq 0$ is satisfied as long as $\rho_{\rm d}(r) > 0$ and $\rho_{\rm d}'(r) < 0$, while $\rho_{\rm d}(r)$ and its mass integral $m_{\rm d}(r)$ remain finite everywhere; these conditions hold for the Einasto profile with $n > 0$.

\subsection{Modified Tolman--Oppenheimer--Volkoff Equations}
\label{ssec:tov}

We adopt the static, spherically symmetric metric ansatz:
\begin{equation}
\label{g}
ds^2 = -e^{2\alpha(r)} c^2 dt^2 + e^{2\beta(r)} dr^2 + r^2 (d\theta^2 + \sin^2\theta \, d\varphi^2).
\end{equation}

The ordinary baryonic matter is treated as a perfect fluid:
\begin{equation}
\label{tm}
T_{\mu\nu}^{\rm (b)} = (\rho c^2 + p) U_\mu U_\nu + p \, g_{\mu\nu},
\end{equation}
with the four-velocity normalized as $U^\mu U_\mu = -1$, giving $U_\mu = (e^{\alpha(r)}, 0, 0, 0)$.

The $G_{tt}$ and $G_{rr}$ components of the Einstein equations yield:
\begin{align}
\frac{e^{-2\beta(r)}}{r^2} \left( 2r\beta'(r) + e^{2\beta(r)} - 1 \right) &= \frac{8\pi G}{c^2} \left( \rho_{\rm d}(r) + \rho(r) \right), \label{G00} \\
\frac{e^{-2\beta(r)}}{r^2} \left( 2r\alpha'(r) - e^{2\beta(r)} + 1 \right) &= \frac{8\pi G}{c^4} \left( p_{r{\rm (d)}}(r) + p(r) \right). \label{G11}
\end{align}

We define the mass functions for the baryonic matter and the dark matter respectively:
\begin{align}
m(r) &= 4\pi \int_0^r \rho(x) \, x^2 \, dx, \label{mb} \\
m_{\rm d}(r) &= 4\pi \int_0^r \rho_{\rm d}(x) \, x^2 \, dx. \label{md}
\end{align}
The total gravitational mass enclosed within radius $r$ is then $m_{\rm all}(r) = m(r) + m_{\rm d}(r)$.

From Eq.~(\ref{G00}), we identify the metric function ($g_{rr}^{-1}$)
\begin{equation}
\label{beta}
e^{-2\beta(r)} = 1 - \frac{2G \, m_{\rm all}(r)}{c^2 r}.
\end{equation}
The total energy-momentum tensor is conserved, $\nabla_\mu (T^{\mu\nu}_{\rm (b)} + T^{\mu\nu}_{\rm (d)}) = 0$. Since the two sectors are independently conserved, $\nabla_\mu T^{\mu\nu}_{\rm (b)} = 0$ gives
\begin{equation}
(\rho(r) c^2 + p(r))\,\alpha'(r) + p'(r) = 0.
\label{hydrostatic}
\end{equation}
Combining this with Eqs.~(\ref{G11}), (\ref{beta}), and (\ref{hydrostatic}) yields the modified TOV equations:
\begin{align}
\frac{dm}{dr} &= 4\pi r^2 \rho(r), \label{TOV1} \\
\frac{dp}{dr} &= -\frac{G}{c^2 r} \, \frac{\left( c^2 m_{\rm all}(r) + 4\pi r^3 p_{\rm all}(r) \right) \left( p(r) + c^2 \rho(r) \right)}{c^2 r - 2G \, m_{\rm all}(r)}. \label{TOV2}
\end{align}
Here $p_{\rm all}(r) = p(r) + p_{r{\rm (d)}}(r)$ is the total radial pressure, including the dark matter contribution $p_{r{\rm (d)}} = -\rho_{\rm d} c^2$.

\section{Numerical Results}
\label{sec:numerical}

Throughout this paper, numerical solutions to the initial value problem are obtained with an adaptive fourth-order Runge--Kutta method. The modified TOV equations are solved from the center at $r=r_{\delta}$ to the surface ($p=\rho=0$) of the star at $r=R$, satisfying the boundary conditions:
\begin{equation}
m(r_{\delta}) = \frac{4}{3}\pi r_{\delta}^3 \rho_c,\quad \rho(r_{\delta}) c^2 = \rho_c c^2,\quad p(r_{\delta}) = p_c,
\end{equation}
where $\rho_c$ and $p_c$ are the central density and pressure of the baryonic matter, respectively, and $r_{\delta}=10^{-10}~[\rm{m}]\ll R$. We have checked that all results are independent of the choice of $r_{\delta}$ provided it is sufficiently small compared to the neutron star radius.
The baryonic matter is described by the BSk19~\cite{Potekhin:2013qqa} and SLy4~\cite{Douchin:2001sv,Haensel:2004nu} equations of state. We probe values of central density between $\rho_c = 2.5\times 10^{17}~[\rm{kg/m}^3]$ and the causal limit $\rho_c = \rho_{\max}$, at which the speed of sound reaches the speed of light, $v_s = c$. The maximum allowed densities for the two equations of state are listed in Tab.~\ref{tab:rhomax}.

\begin{table}[htbp]
\renewcommand\arraystretch{1.5}
\captionsetup{justification=raggedright}
\caption{Maximum allowed density from causality-constrained equations of state.\label{tab:rhomax}}
\begin{tabular}{c@{\hspace{8mm}}c@{\hspace{8mm}}c}
\hline
EOS & BSk19 & SLy4 \\
\hline
$\rho_{\max}~[\rm{kg/m}^3]$ & $3.3814\times 10^{18}$ & $3.0075\times 10^{18}$ \\
\hline
\end{tabular}
\end{table}

\subsection{Dark Matter Halo Effects on Neutron Star Structure}

The dark matter halo density is described by the Einasto profile~\cite{Einasto:1965czb}:
$\rho_{\rm d}(r) = \rho_0 \exp[-(r/h)^{1/n}]$, which contains three adjustable parameters:
\begin{enumerate}
\item \textbf{Central density $\rho_0$} controls the overall amplitude of the halo. Larger $\rho_0$ increases the dark matter density at all radii, raising the total enclosed DM mass $m_{\rm d}(r)$.
\item \textbf{Scale radius $h$} sets the radial extent over which the density decays. At fixed $\rho_0$ and $n$, larger $h$ extends the halo to greater distances, broadening the region of significant DM contribution.
\item \textbf{Einasto index $n$} governs the steepness of the fall-off. Smaller $n$ produces a steeper profile that concentrates DM near the center; larger $n$ yields a shallower, more extended distribution, which increases the enclosed mass $m_{\rm d}(r)$ at large radii for a given $\rho_0$ and $h$.
\end{enumerate}
These parameters enter Eq.~(\ref{TOV2}) through the DM mass term $m_{\rm d}(r)$. Their combined effect profoundly influences the neutron star structure.

\begin{figure*}[t]
\centering
\includegraphics[width=\textwidth]{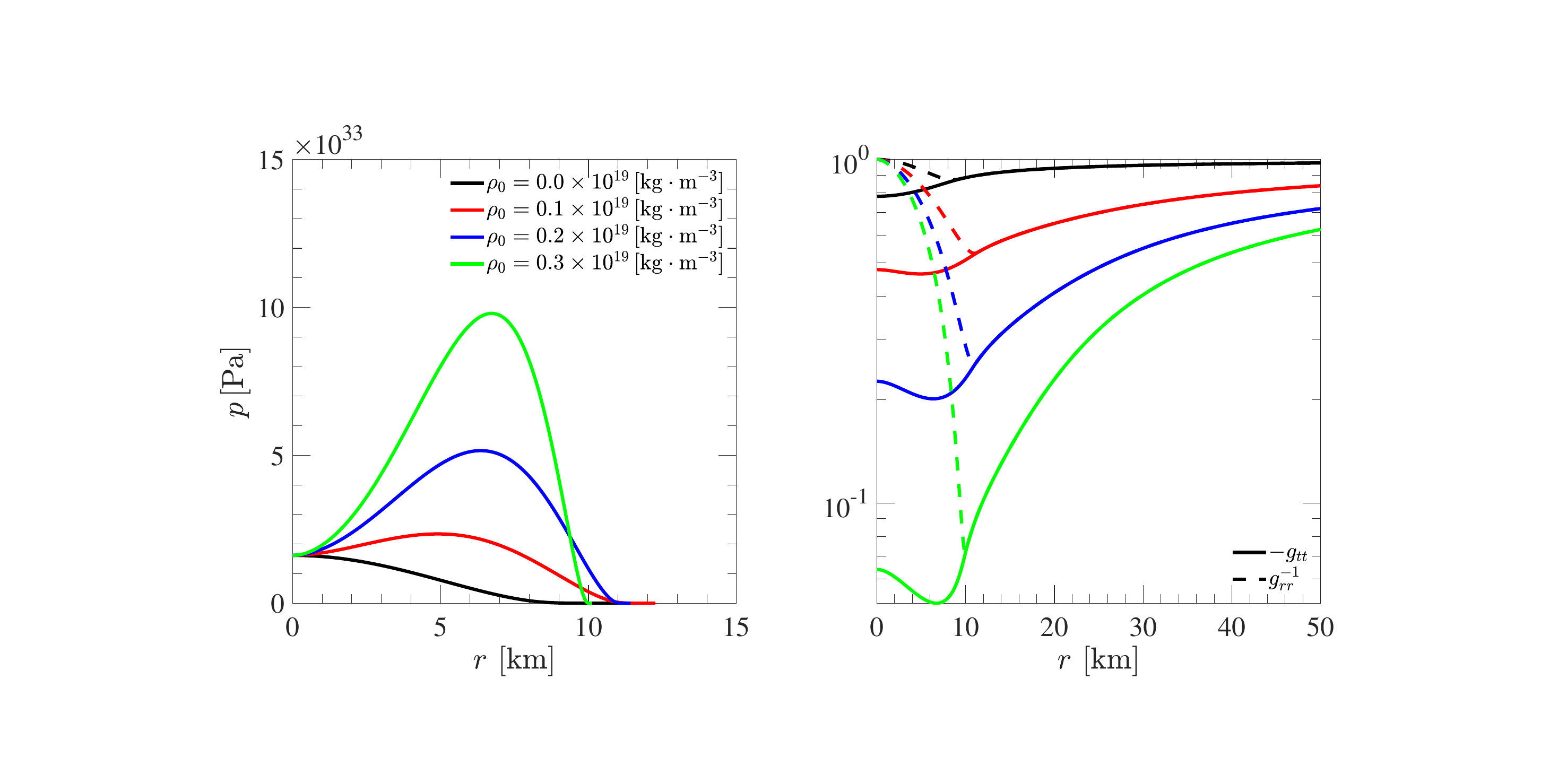}
\caption{Radial pressure profiles (left) and metric functions $-g_{tt}$ (solid) and $g_{rr}^{-1}$ (dashed, right) for different values of $\rho_0$, with $h = 5~[\rm{km}]$, $n = 1$, and $\rho_c = 0.5\times 10^{18}~[\rm{kg/m}^3]$ fixed, computed with the BSk19 equation of state.\label{fig:rho_scan}}
\end{figure*}

\begin{figure*}[t]
\centering
\includegraphics[width=\textwidth]{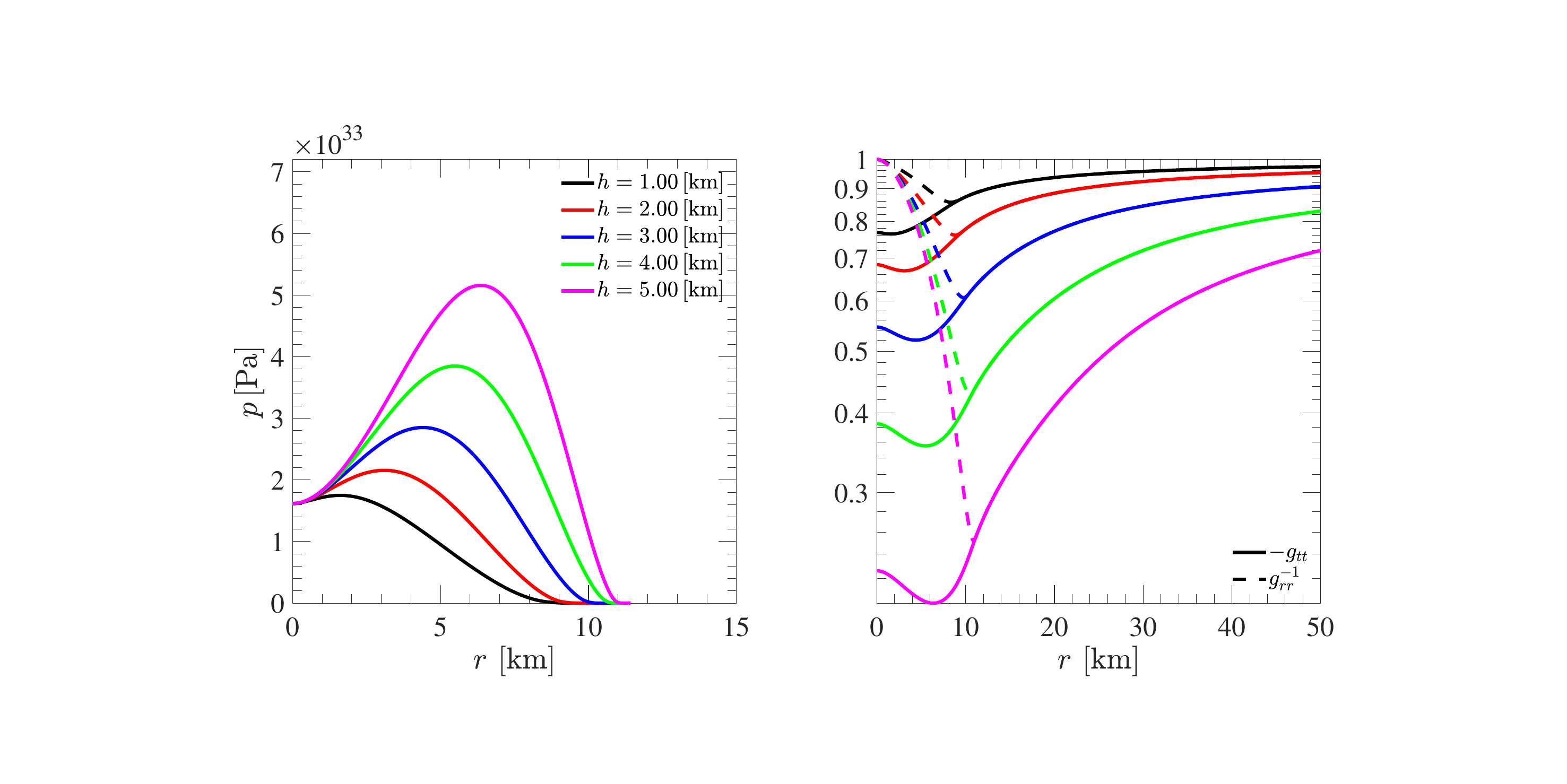}
\caption{Radial pressure profiles (left) and metric functions (right) for different values of $h$, with $\rho_0 = 0.2 \times 10^{19}~[\rm{kg/m}^3]$, $n = 1$, and $\rho_c = 0.5\times 10^{18}~[\rm{kg/m}^3]$ fixed, computed with the BSk19 equation of state.\label{fig:h_scan}}
\end{figure*}

\begin{figure*}[t]
\centering
\includegraphics[width=\textwidth]{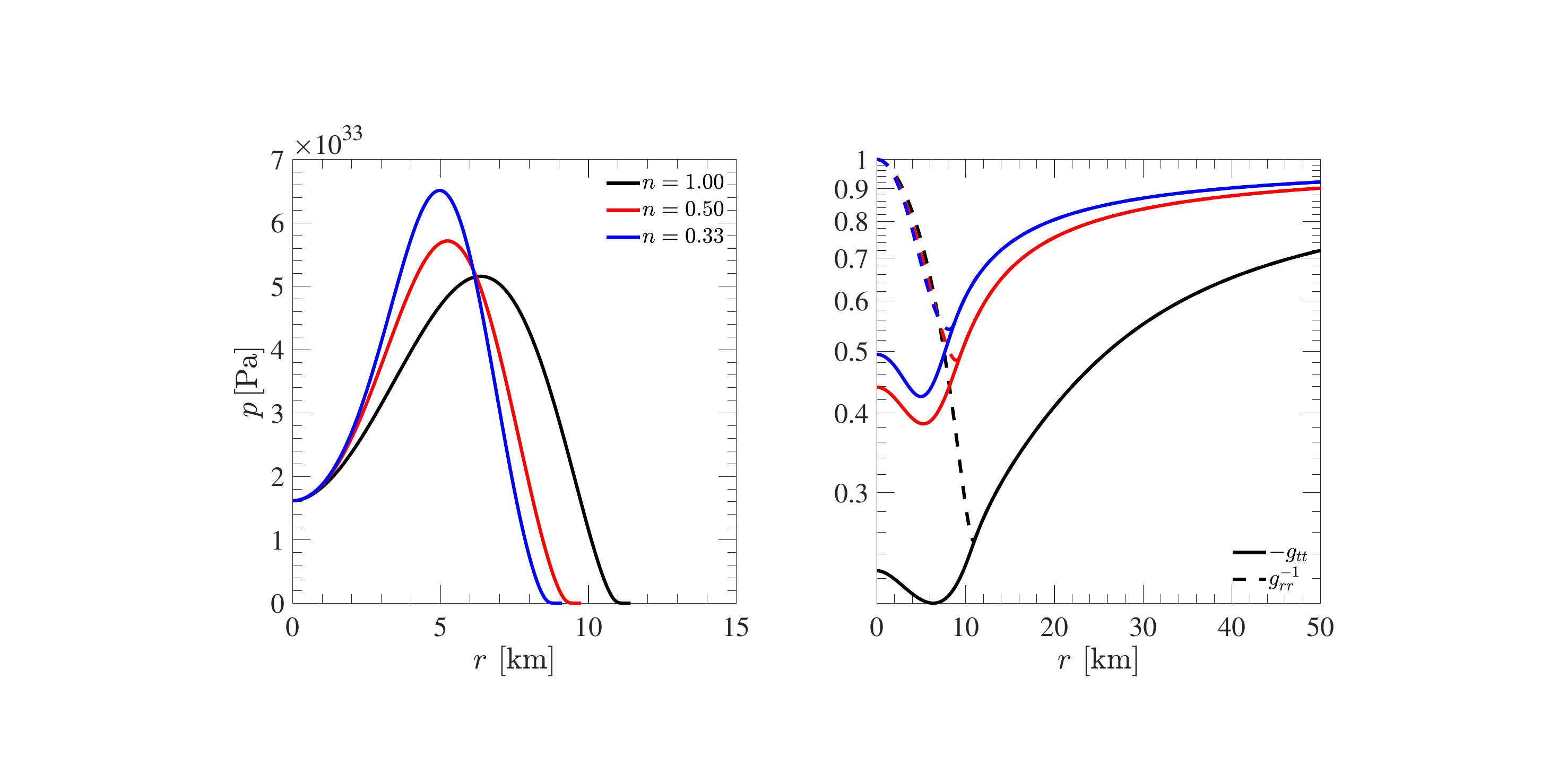}
\caption{Radial pressure profiles (left) and metric functions (right) for different values of $n$, with $\rho_0 = 0.2 \times 10^{19}~[\rm{kg/m}^3]$, $h = 5~[\rm{km}]$, and $\rho_c = 0.5\times 10^{18}~[\rm{kg/m}^3]$ fixed, computed with the BSk19 equation of state.\label{fig:n_scan}}
\end{figure*}

The radial pressure profiles were obtained by solving the modified TOV equations~(\ref{TOV1})--(\ref{TOV2}) with the BSk19 equation of state.
Figs.~\ref{fig:rho_scan}--\ref{fig:n_scan}, taking BSk19 as a representative example, illustrate the structural response of the neutron star to variations in the Einasto dark matter halo parameters, with one parameter scanned while the other two and the neutron star central density $\rho_c = 0.5\times 10^{18}~[\rm{kg/m}^3]$ are held fixed.
In each figure, the left panel displays the radial pressure $p(r)$, and the right panel shows the metric functions $-g_{tt}(r)$ (solid curves) and $g_{rr}^{-1}(r)$ (dashed curves).

The effect of the dark matter halo central density $\rho_0$ on the neutron star structure is shown in Fig.~\ref{fig:rho_scan}, which presents the radial pressure profiles and metric functions for several values of $\rho_0$ with $h = 5~[\rm{km}]$, $n = 1$, and $\rho_c = 0.5\times 10^{18}~[\rm{kg/m}^3]$ fixed.
The $\rho_0 = 0$ case corresponds to an ordinary neutron star without dark matter.
As $\rho_0$ increases from zero, the radius first expands and then contracts (Tab.~\ref{tab:rho_scan}), reflecting a competition between the negative radial pressure $p_{r{\rm (d)}} = -\rho_{\rm d} c^2$ (Eq.~\ref{pr}) and the gravitational attraction of the dark matter halo.
At low $\rho_0$, the negative pressure dominates and $R$ increases; at larger $\rho_0$, the gravitational attraction overwhelms the negative pressure, and the boundary contracts inward (see Tab.~\ref{tab:rho_scan} for the numerical values).
Across this parameter range, a dense, high-pressure region of baryonic matter develops in the stellar interior with its maximum pressure growing with $\rho_0$, producing a ``filled hard candy''-like structure analogous to that reported in Ref.~\cite{Tan:2025jcg}. The minima of both $-g_{tt}$ and $g_{rr}^{-1}$ decrease with increasing $\rho_0$.

\begin{table}[htbp]
\renewcommand\arraystretch{1.5}
\captionsetup{justification=raggedright}
\caption{Neutron star radius $R$ as a function of the dark matter halo central density $\rho_0$, with $h=5~[\rm{km}]$, $n=1$, and $\rho_c=0.5\times10^{18}~[\rm{kg/m}^3]$ fixed.\label{tab:rho_scan}}
\begin{tabular}{c|ccccc}
\hline\hline
$\rho_0$ [$10^{19}~\rm{kg/m}^3$] & 0.0 & 0.1 & 0.2 & 0.3 \\
\hline
$R$ [km] & 11.8748 & 12.2568 & 11.4243 & 10.1228 \\
\hline
\end{tabular}
\end{table}

The effect of the dark matter halo scale radius $h$ on the neutron star structure is shown in Fig.~\ref{fig:h_scan}, which presents the radial pressure profiles and metric functions for several values of $h$ with $\rho_0 = 0.2 \times 10^{19}~[\rm{kg/m}^3]$, $n = 1$, and $\rho_c = 0.5\times 10^{18}~[\rm{kg/m}^3]$ fixed.
As $h$ increases, the radius first contracts and then expands (Tab.~\ref{tab:h_scan}).
At small $h$, the dark matter is concentrated near the center, gravitational compression dominates, and $R$ decreases; at larger $h$, the halo broadens and the negative pressure effect becomes more extended, causing $R$ to increase (see Tab.~\ref{tab:h_scan} for the numerical values).
Across this parameter range, a dense, high-pressure region of baryonic matter develops in the stellar interior with its maximum pressure growing with $h$, and the minima of both $-g_{tt}$ and $g_{rr}^{-1}$ decrease with increasing $h$.

\begin{table}[htbp]
\renewcommand\arraystretch{1.5}
\captionsetup{justification=raggedright}
\caption{Neutron star radius $R$ as a function of the dark matter halo scale radius $h$, with $\rho_0=0.2\times10^{19}~[\rm{kg/m}^3]$, $n=1$, and $\rho_c=0.5\times10^{18}~[\rm{kg/m}^3]$ fixed.\label{tab:h_scan}}
\begin{tabular}{c|cccccc}
\hline\hline
$h$ [km] & 1.0 & 2.0 & 3.0 & 4.0 & 5.0 \\
\hline
$R$ [km] & 11.2881 & 10.6732 & 11.0027 & 11.3498 & 11.4243 \\
\hline
\end{tabular}
\end{table}

The effect of the Einasto index $n$ on the neutron star structure is shown in Fig.~\ref{fig:n_scan}, which presents the radial pressure profiles and metric functions for several values of $n$ with $\rho_0 = 0.2 \times 10^{19}~[\rm{kg/m}^3]$, $h = 5~[\rm{km}]$, and $\rho_c = 0.5\times 10^{18}~[\rm{kg/m}^3]$ fixed.
As $n$ increases, the radius grows monotonically (Tab.~\ref{tab:n_scan}): a shallower profile distributes the dark matter over a larger volume, which reduces the enclosed gravitational mass near the center and extends the negative radial pressure over a broader region, both weakening the compression.
Across this parameter range, a dense, high-pressure region of baryonic matter develops in the stellar interior with its maximum pressure decreasing with $n$, and the minima of both $-g_{tt}$ and $g_{rr}^{-1}$ decrease with increasing $n$.

\begin{table}[htbp]
\renewcommand\arraystretch{1.5}
\captionsetup{justification=raggedright}
\caption{Neutron star radius $R$ as a function of the dark matter halo Einasto index $n$, with $\rho_0=0.2\times10^{19}~[\rm{kg/m}^3]$, $h=5~[\rm{km}]$, and $\rho_c=0.5\times10^{18}~[\rm{kg/m}^3]$ fixed.\label{tab:n_scan}}
\begin{tabular}{c|ccc}
\hline\hline
$n$ & 1/3 & 1/2 & 1 \\
\hline
$R$ [km] & 9.1108 & 9.7479 & 11.4243 \\
\hline
\end{tabular}
\end{table}

We note that the three Einasto parameters cannot be increased without limit.
The baryonic equation of state has a maximum density and pressure beyond which causality is violated; when the DM parameters become too large, the additional gravitational compression forces the pressure of the baryonic matter outside the equation of state validity range, and no static solution exists.
This places a physical upper bound on the viable parameter space.

\subsection{Horizon Formation: Neutron Stars in Black Holes}

The neutron star resides entirely within the horizon, a ``neutron star in a black hole.'' What distinguishes this configuration from an ordinary black hole is that the event horizon is not produced by the gravitational collapse of the central object. The gravitational field is provided jointly by the neutron star and the dark matter halo coexisting with it, and the dark matter halo plays an important role in the formation of the horizon, while the star itself persists as a regular, non-singular structure in the interior. This distinctive geometry is illustrated schematically in Fig.~\ref{fig:schematic}.

\begin{figure*}[t]
\centering
\includegraphics[width=\textwidth]{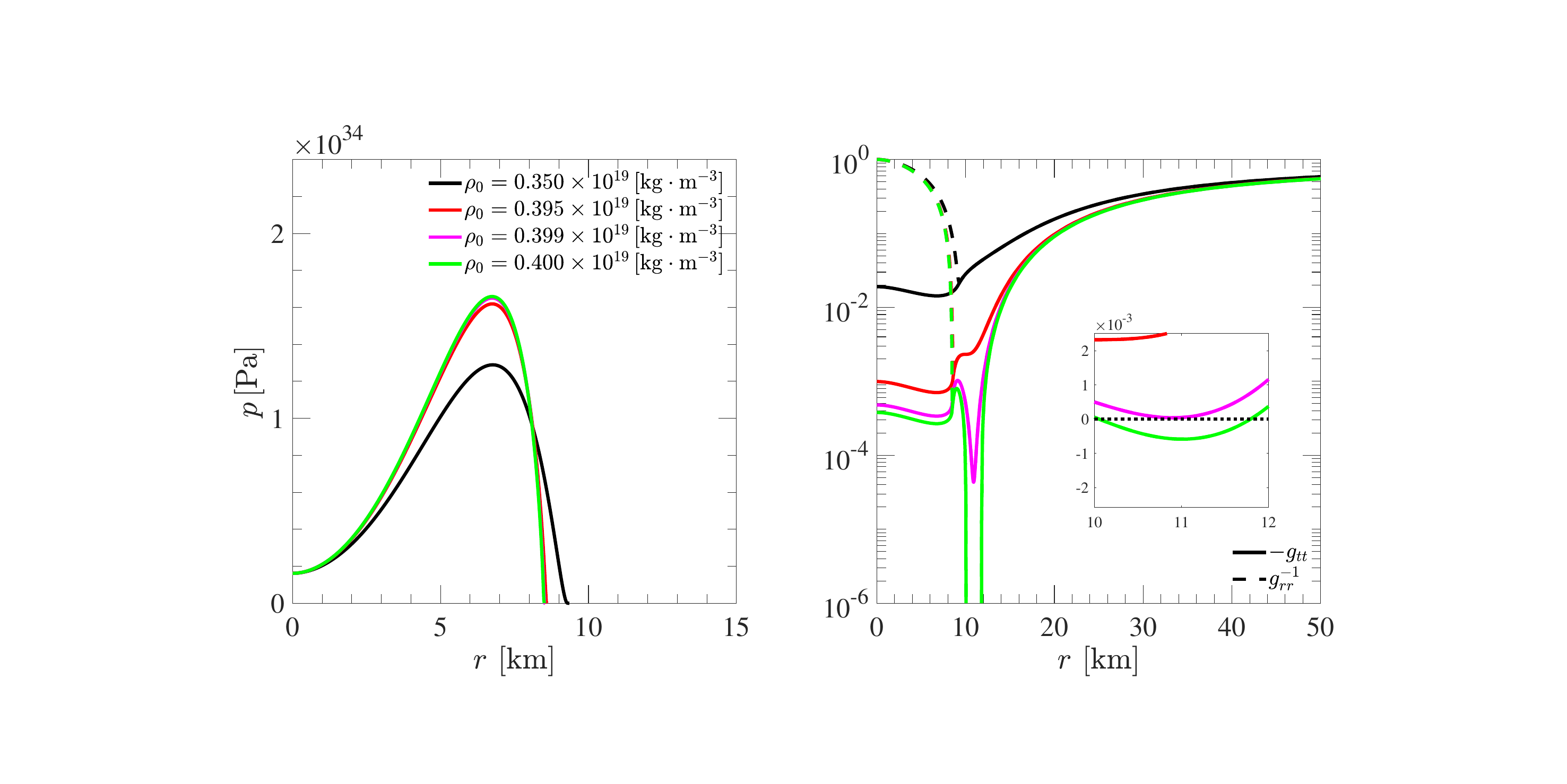}
\caption{Approach to horizon formation as the central dark matter density $\rho_0$ increases, with $h=5~[\rm{km}]$, $n=1$, and $\rho_c=0.5\times10^{18}\,[\rm{kg/m^3}]$, computed with the BSk19 equation of state. \textit{Left:} radial pressure profiles. \textit{Right:} metric functions $-g_{tt}$ (solid) and $g_{rr}^{-1}$ (dashed). The inset provides a magnified view of the horizon crossing, where the curve for $\rho_0=0.40\times10^{19}$ drops below zero.\label{fig:horizon_transition}}
\end{figure*}

We first encountered this configuration using the same parameters as Fig.~\ref{fig:rho_scan} ($h=5~[\rm{km}]$, $n=1$, $\rho_c=0.5\times10^{18}~[\rm{kg/m}^3]$): as $\rho_0$ is increased beyond the values shown there, an event horizon gradually emerges in the space outside the stellar surface, a process tracked in detail by Fig.~\ref{fig:horizon_transition}, while the neutron star solution continues to exist. At $\rho_0=0.35\times10^{19}~[\rm{kg/m}^3]$, $g_{rr}^{-1}$ remains positive everywhere with a single local minimum, corresponding to an ordinary DANS. As $\rho_0$ increases, a second local minimum of $g_{rr}^{-1}$ appears in the exterior region. This second minimum drops below the interior minimum and continues to decrease, until at $\rho_0=0.40\times10^{19}~[\rm{kg/m}^3]$ it becomes negative ($\min(g_{rr}^{-1})=-5.8152\times 10^{-4}$) in a narrow radial shell $10.06~[\rm{km}]<r<11.79~[\rm{km}]$ outside the stellar surface ($R=8.51~[\rm{km}]$), signaling the formation of an event horizon. The neutron star interior remains regular, with $g_{rr}^{-1}>0$ throughout $r\leq R$. When $\rho_0$ exceeds $0.41\times10^{19}~[\rm{kg/m}^3]$, the required pressure of the baryonic matter exceeds the equation-of-state upper limit, hydrostatic equilibrium is destroyed, and no static solution exists.

\begin{figure}[t]
\centering
\includegraphics[width=\columnwidth]{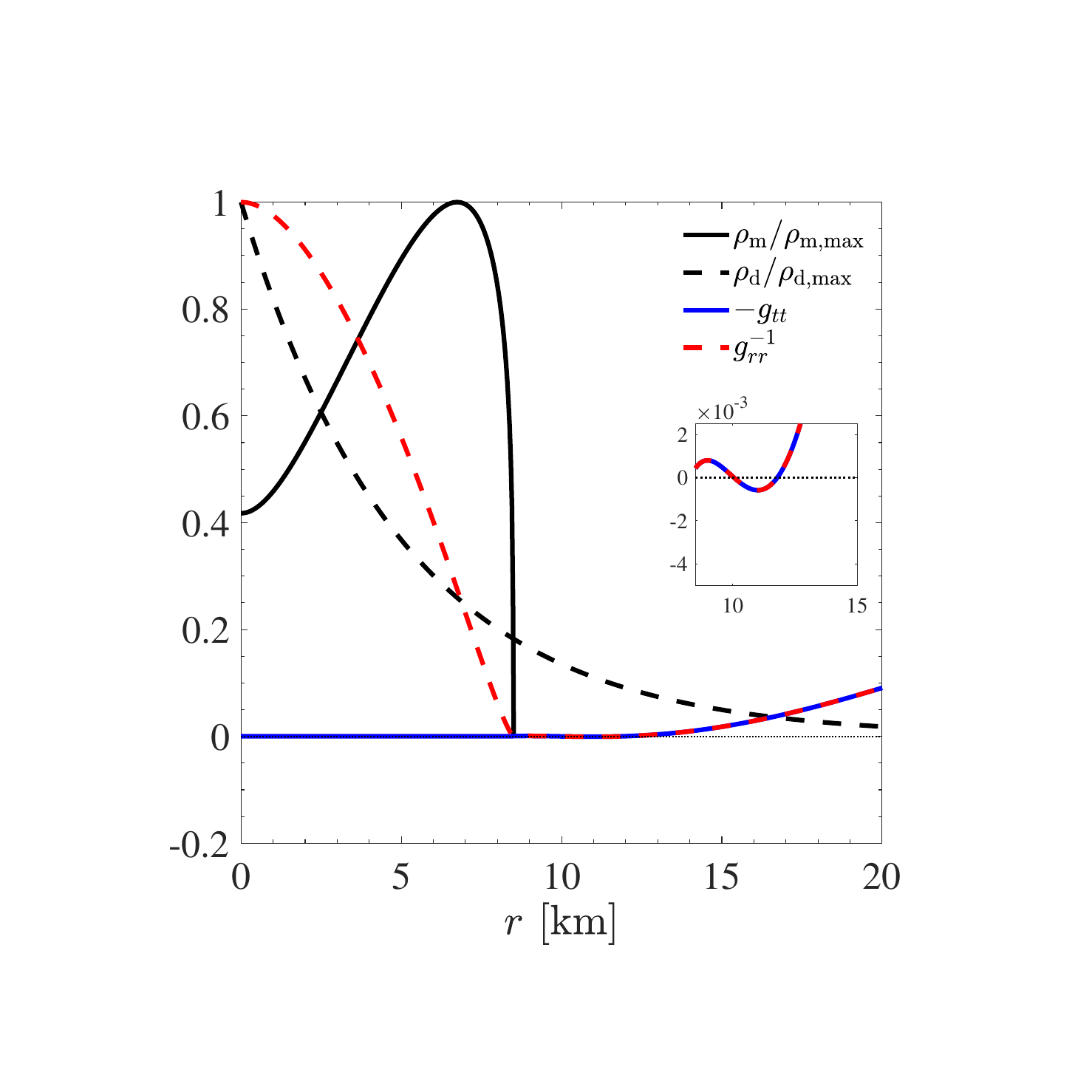}
\caption{Radial profiles of the normalized matter densities $\rho_{\rm b}/\rho_{\rm b,max}$ (solid black) and $\rho_{\rm d}/\rho_{\rm d,max}$ (dashed black), together with the metric functions $-g_{tt}$ (solid blue) and $g_{rr}^{-1}$ (dashed red), for $\rho_0 = 0.40\times10^{19}\,[\rm{kg/m^3}]$, $h=5~[\rm{km}]$, $n=1$, and $\rho_c = 0.5\times10^{18}\,[\rm{kg/m^3}]$, computed with the BSk19 equation of state. \label{fig:horizon_n1}}
\end{figure}

Fig.~\ref{fig:horizon_n1} provides crucial insight into this behavior. The dark matter density $\rho_{\rm d}$ (dashed black) extends significantly beyond the neutron star surface ($R=8.51~[\rm{km}]$, dotted line), whereas the baryonic matter density $\rho_{\rm b}$ (solid black) terminates at the stellar boundary. The neutron star mass $m(r)$ ceases to grow beyond $R$, and the exterior evolution of the metric is governed solely by the increase of $r$ and the mass contribution of the dark matter halo $m_{\rm d}(r)$. Correspondingly, the minimum of $g_{rr}^{-1}$ (dashed red) lies in the region where only dark matter density is present. The growth of the dark matter density outside the stellar surface thus gives rise to the exterior local minimum of $g_{rr}^{-1}$, and the dark matter halo plays an important role in the formation of the event horizon.

\begin{figure*}[t]
\centering
\subfloat[$h$--$\rho_0$ ($n=1.0$).]{\includegraphics[width=0.355\textwidth]{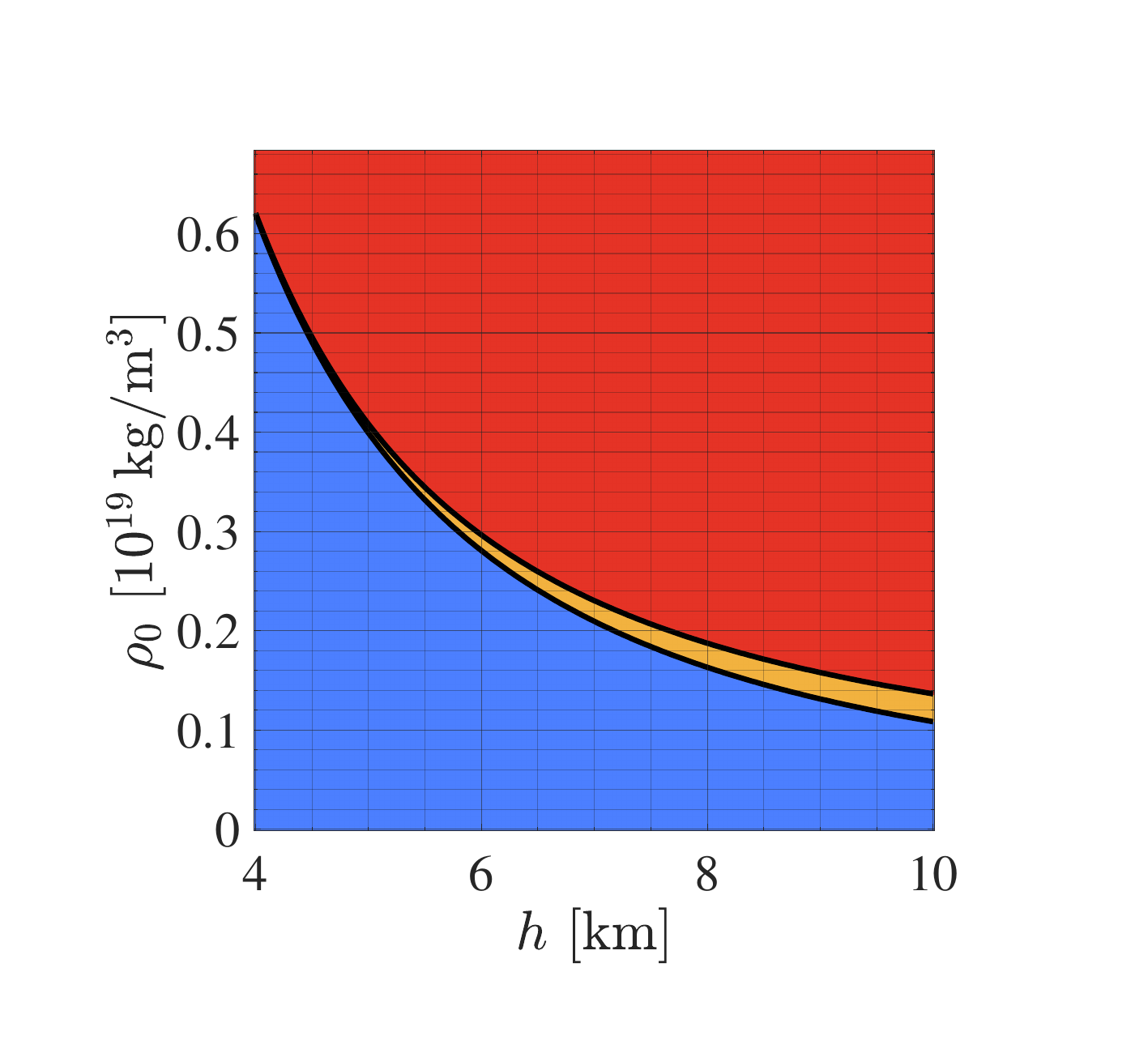}\label{fig:phase_h_rho}}\hspace{-12mm}
\subfloat[$n$--$\rho_0$ ($h=5$ km).]{\includegraphics[width=0.355\textwidth]{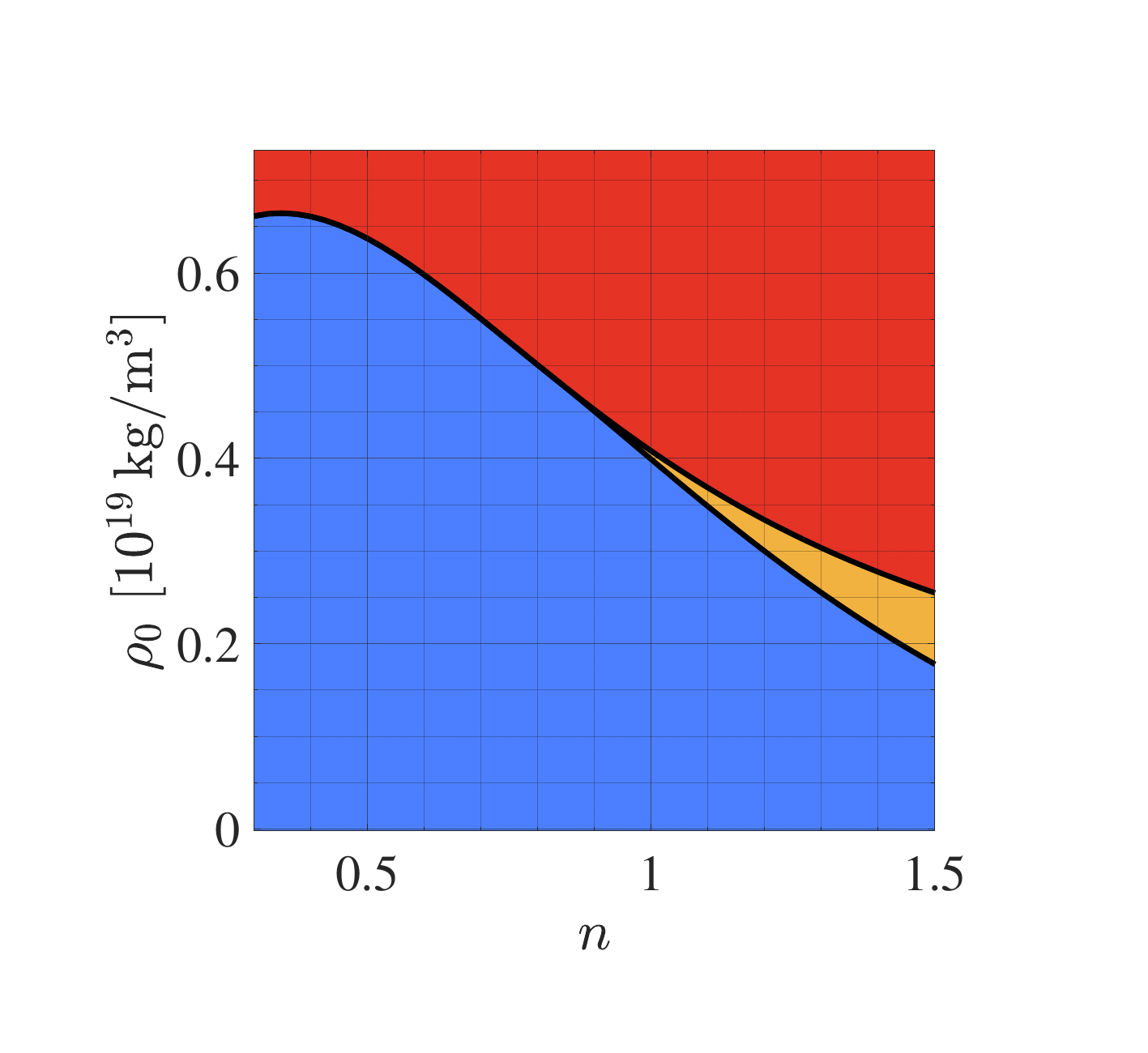}\label{fig:phase_n_rho}}\hspace{-12mm}
\subfloat[$n$--$h$ ($\rho_0=0.2\times10^{19}$~kg/m$^3$).]{\includegraphics[width=0.355\textwidth]{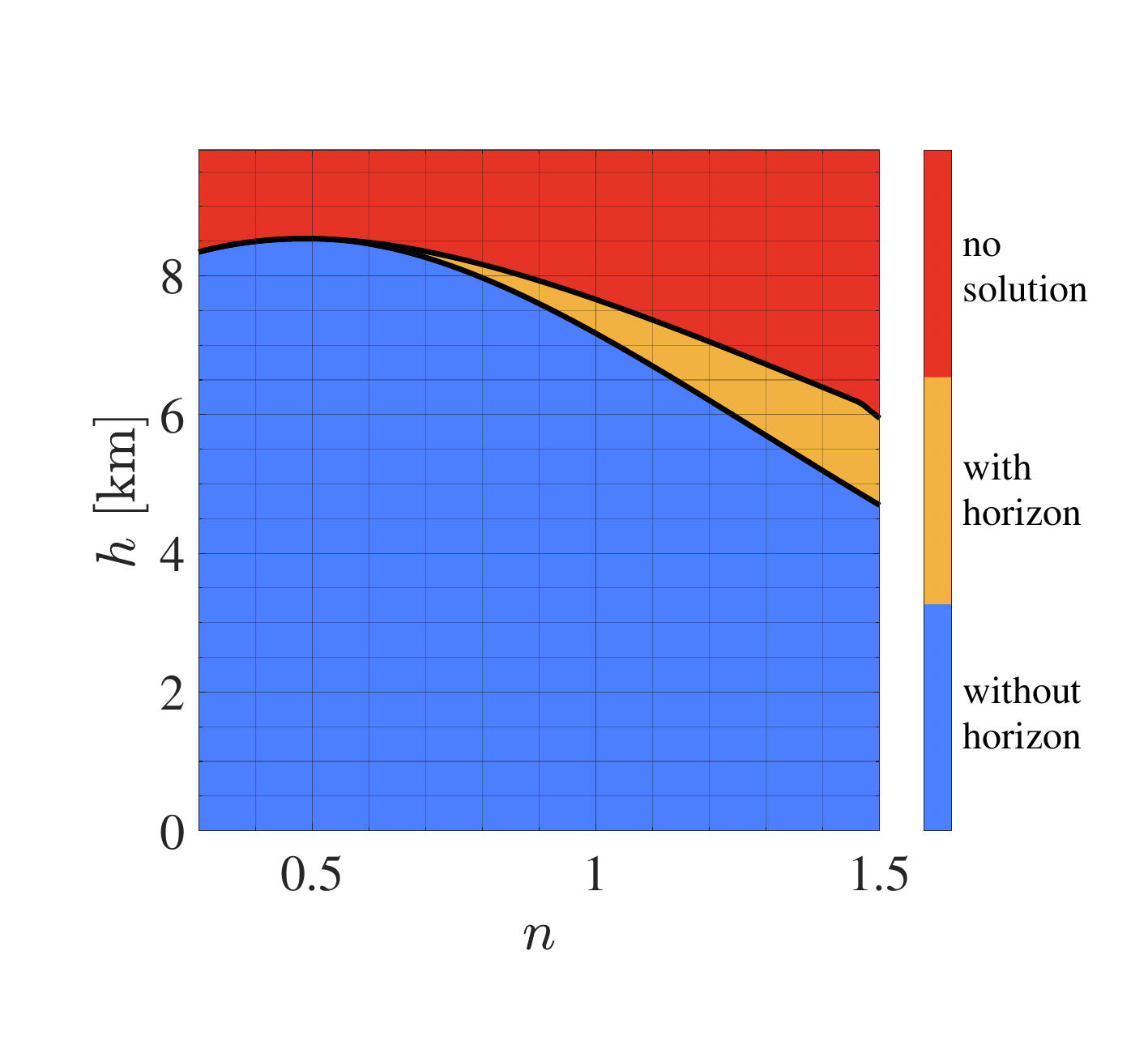}\label{fig:phase_n_h}}
\caption{Phase diagrams of the three solution regimes in the Einasto parameter space, computed with the BSk19 equation of state. Blue: ordinary DANSs (no horizon). Orange: neutron star in a black hole (horizon present). Red: no static solution (collapse). Solid lines mark the horizon and collapse boundaries.\label{fig:phase_diagrams}}
\end{figure*}

\begin{figure}[t]
\centering
\includegraphics[width=\columnwidth]{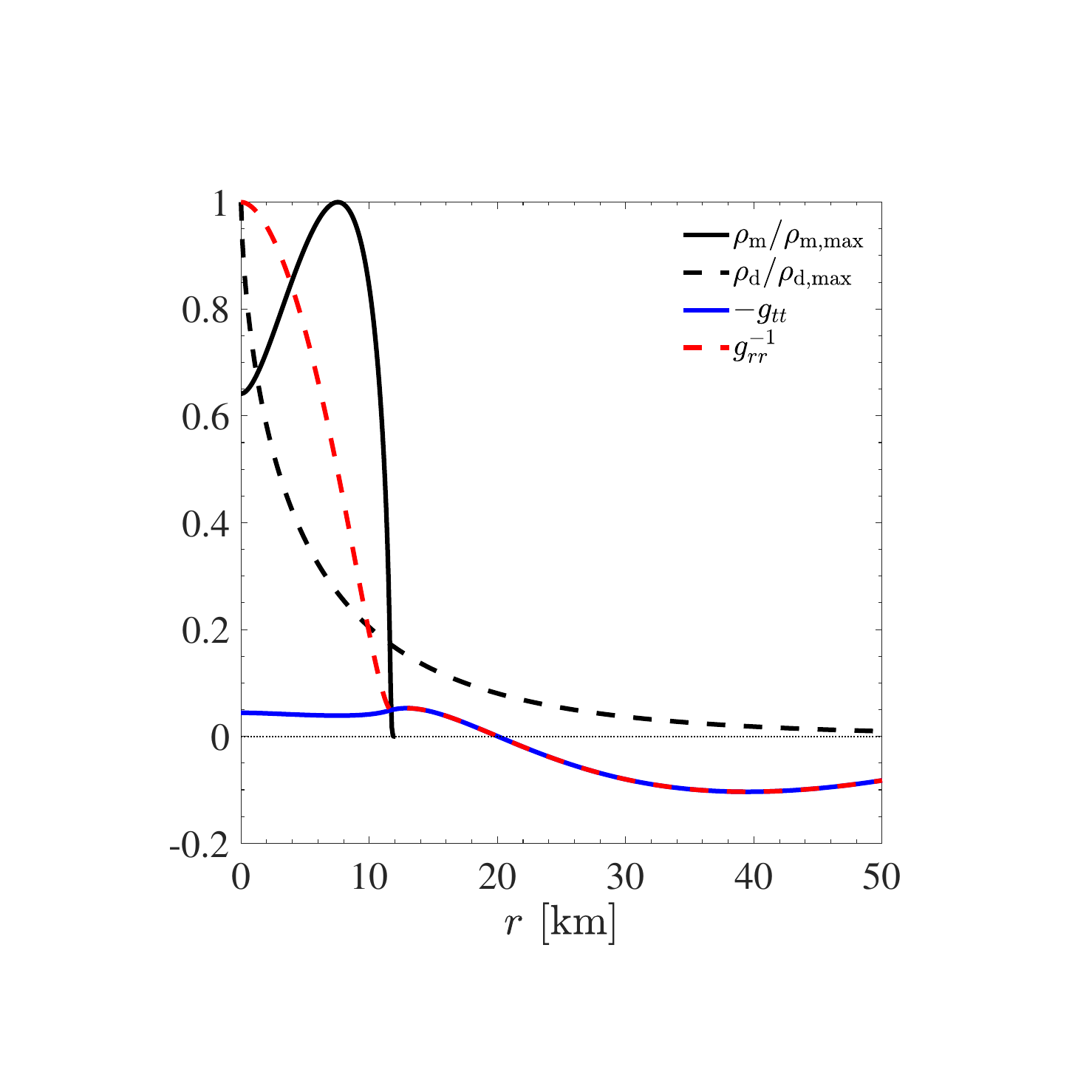}
\caption{Radial profiles of the normalized matter densities $\rho_{\rm b}/\rho_{\rm b,max}$ (solid black) and $\rho_{\rm d}/\rho_{\rm d,max}$ (dashed black), together with the metric functions $-g_{tt}$ (solid blue) and $g_{rr}^{-1}$ (dashed red), for $n=1.5$, $\rho_0 = 0.20\times10^{19}\,[\rm{kg/m^3}]$, $h=5~[\rm{km}]$, and $\rho_c = 0.5\times10^{18}\,[\rm{kg/m^3}]$, computed with the BSk19 equation of state.\label{fig:horizon_n15}}
\end{figure}

This configuration occupies a relatively broad parameter space, as illustrated by the three phase diagrams in Fig.~\ref{fig:phase_diagrams}. These diagrams, computed with the BSk19 equation of state, map the three solution regimes---ordinary DANSs, neutron star in a black hole, and no static solution---in three orthogonal slices of the Einasto parameter space (Figs.~\ref{fig:phase_h_rho}, \ref{fig:phase_n_rho}, and \ref{fig:phase_n_h}: $h$--$\rho_0$ at $n=1$, $n$--$\rho_0$ at $h=5~[\rm{km}]$, and $n$--$h$ at $\rho_0=0.2\times10^{19}~[\rm{kg/m}^3]$, respectively). Although the full three-dimensional parameter space would be considerably more intricate, these two-dimensional slices through it capture a representative portion and reveal a general trend: provided a static neutron star solution exists, variations of the Einasto parameters that increase the total dark matter halo mass tend to favor the formation of this configuration. As an illustrative example, fixing $\rho_c = 0.5\times10^{18}~[\rm{kg/m}^3]$ and increasing the Einasto index to $n=1.5$ with $h=5~[\rm{km}]$, the dark matter central density must be reduced to $\rho_0 = 0.2\times10^{19}~[\rm{kg/m}^3]$ in order to maintain a static neutron star solution. As shown in Fig.~\ref{fig:horizon_n15}, this parameter set likewise produces a neutron star in a black hole, with a wider and deeper negative region of $g_{rr}^{-1}$ than the $n=1$ case. The shallower Einasto profile distributes dark matter over a larger volume, producing a more extended region where $g_{rr}^{-1}<0$. One sees that larger $n$ lowers the critical $\rho_0$ required to form this configuration and broadens the horizon shell.

\subsection{Mass-Radius Relation}

The mass-radius ($M$--$R$) relation is a fundamental issue in neutron star research and provides the most readily accessible observational information. To quantify how the dark matter halo modifies this relation, we construct $M$--$R$ curves by scanning the central baryonic matter density $\rho_c$ for a range of dark matter central densities $\rho_0$, using both BSk19 and SLy4. All mass values are evaluated at the neutron star radius $R$: $M_{\rm NS}=m(R)$, $M_{\rm DM}=m_{\rm d}(R)$, and $M_{\rm all}=m_{\rm all}(R)=M_{\rm NS}+M_{\rm DM}$.

\begin{figure*}[t]
\centering
\subfloat[Total gravitational mass $M_{\rm all}=M_{\rm NS}+M_{\rm DM}$.]{\includegraphics[width=0.48\textwidth]{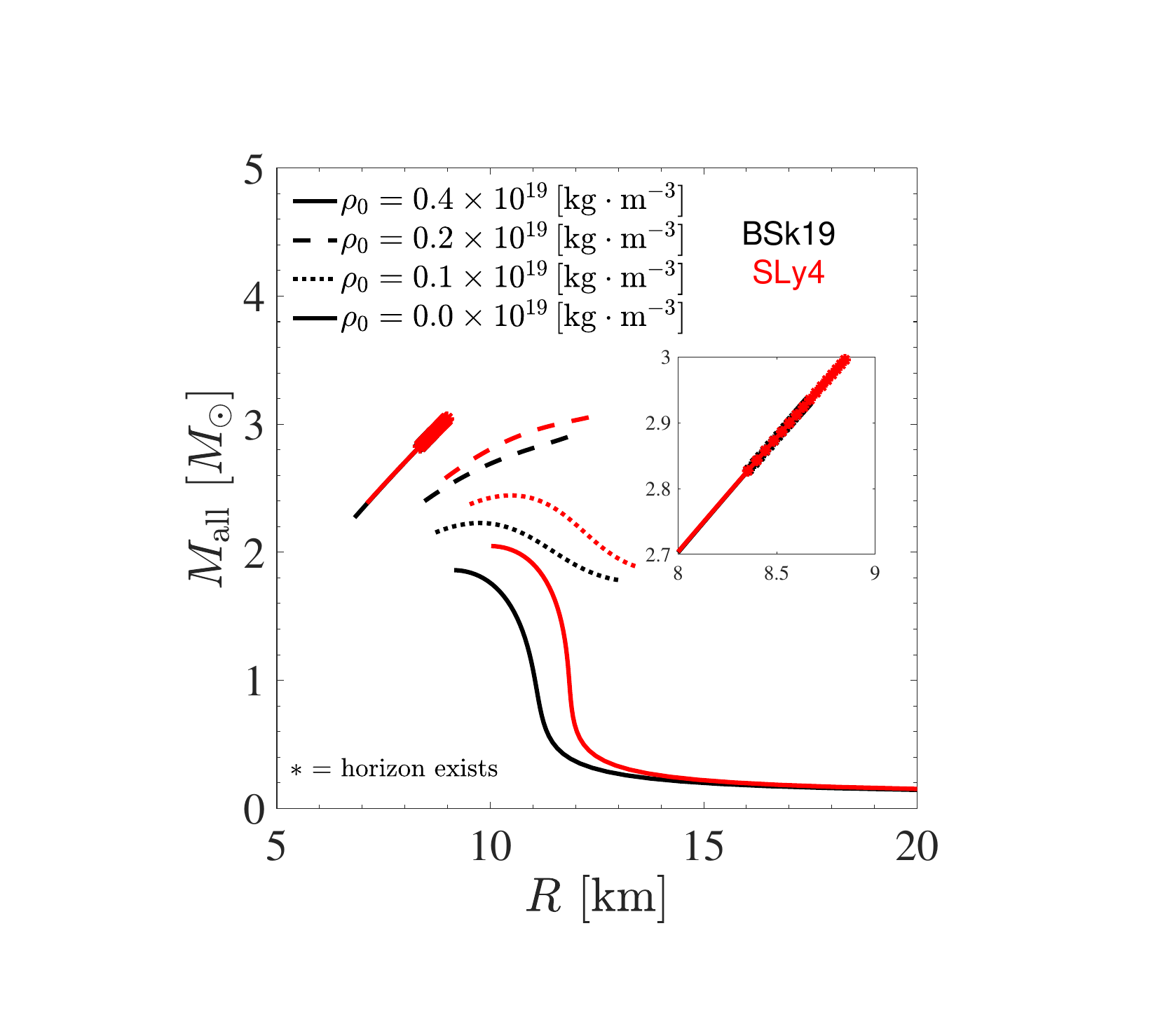}\label{fig:mr_total}}
\hfill
\subfloat[Neutron star mass $M_{\rm NS}$ alone.]{\includegraphics[width=0.48\textwidth]{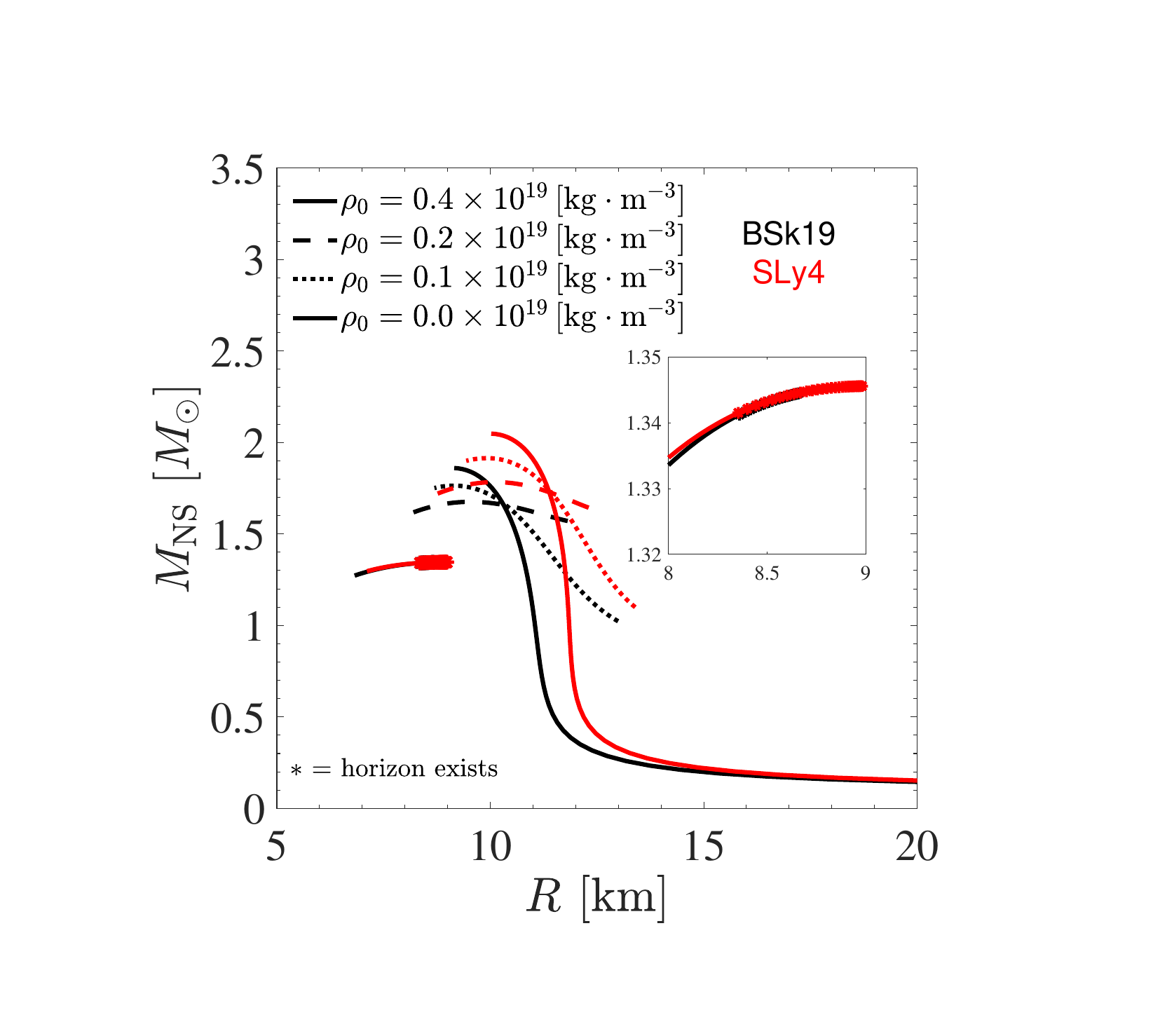}\label{fig:mr_ns}}
\caption{Mass-radius relations for $h=5~[\rm{km}]$ and $n=1$. Solid (BSk19) and dashed (SLy4) curves correspond to different dark matter central densities $\rho_0$. Asterisks mark configurations for which $\min(g_{rr}^{-1})<0$, indicating the existence of a horizon.}
\label{fig:mr_combined}
\end{figure*}

\begin{figure}[t]
\centering
\includegraphics[width=\columnwidth]{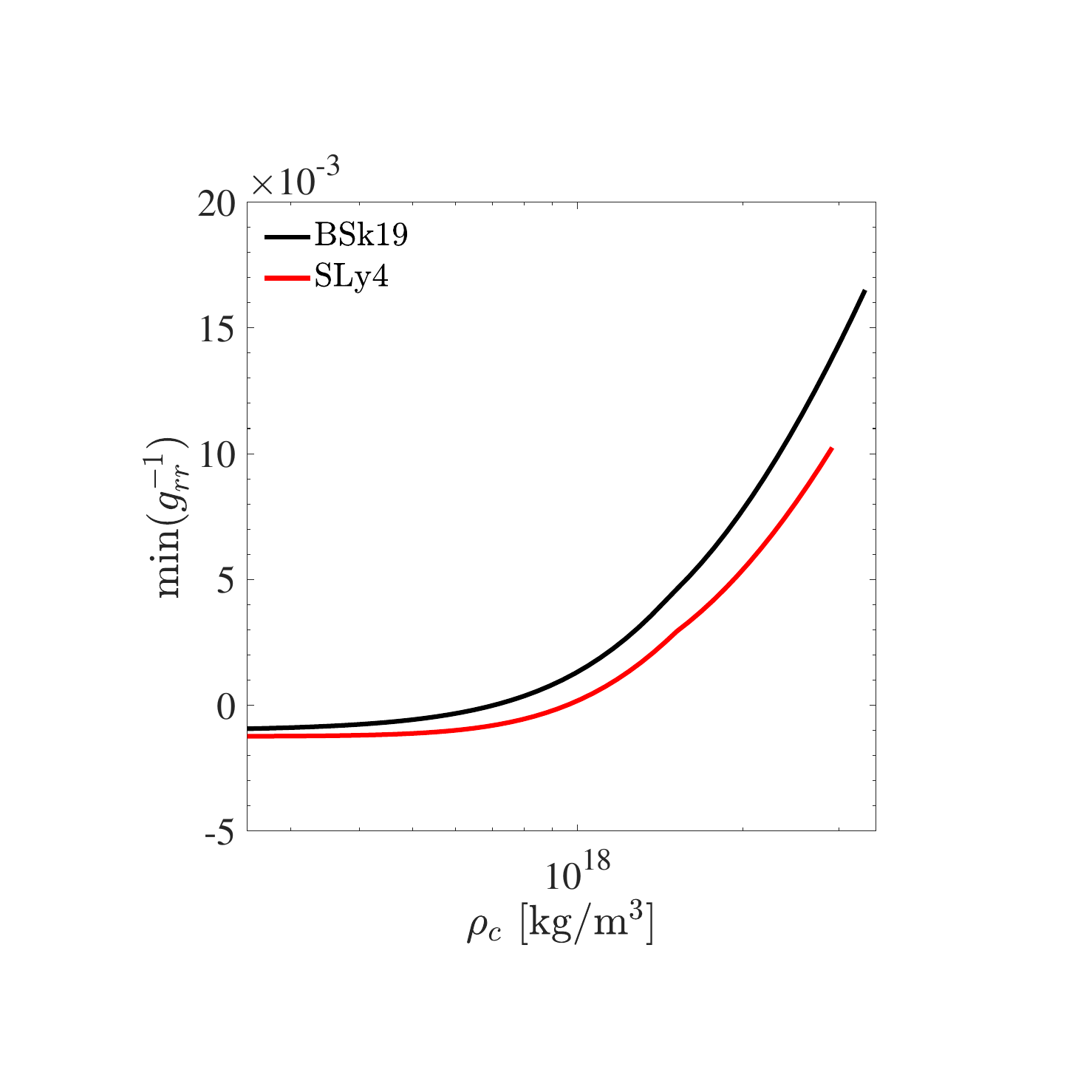}
\caption{Global minimum of $g_{rr}^{-1}$ as a function of the central baryonic matter density $\rho_c$, for $\rho_0=0.40\times10^{19}\,[\rm{kg/m^3}]$, $h=5~[\rm{km}]$, $n=1$.\label{fig:fmin_ed}}
\end{figure}

Fig.~\ref{fig:mr_total} presents the total gravitational mass $M_{\rm all}$ as a function of the neutron star radius $R$, with $h=5~[\rm{km}]$ and $n=1$. $M_{\rm all}$ increases with $\rho_0$, a direct consequence of the growing dark matter contribution to the enclosed mass. Configurations with $\min(g_{rr}^{-1})<0$, i.e., an event horizon has formed, are marked with asterisks; these appear on the right side of the neutron star sequence, at lower central densities $\rho_c$. This occurs because, as seen from Fig.~\ref{fig:mr_ns}, the presence of dark matter significantly alters the neutron star structure: the $M_{\rm NS}$--$R$ relation is shifted so that configurations at lower $\rho_c$ acquire larger neutron star masses, making them more susceptible to horizon formation. This trend is clearly visible in Fig.~\ref{fig:fmin_ed}, where for $\rho_0=0.40\times10^{19}~[\rm{kg/m}^3]$, $\min(g_{rr}^{-1})$ increases with $\rho_c$ and crosses zero at $\rho_c=0.71\times10^{18}~[\rm{kg/m}^3]$ for BSk19 and $\rho_c=0.97\times10^{18}~[\rm{kg/m}^3]$ for SLy4. The neutron-star-in-a-black-hole configuration appears in both BSk19 and SLy4, indicating that it does not depend on a specific choice of the equation of state.

\begin{figure*}[t]
\centering
\subfloat[Total gravitational mass $M_{\rm all}=M_{\rm NS}+M_{\rm DM}$.]{\includegraphics[width=0.48\textwidth]{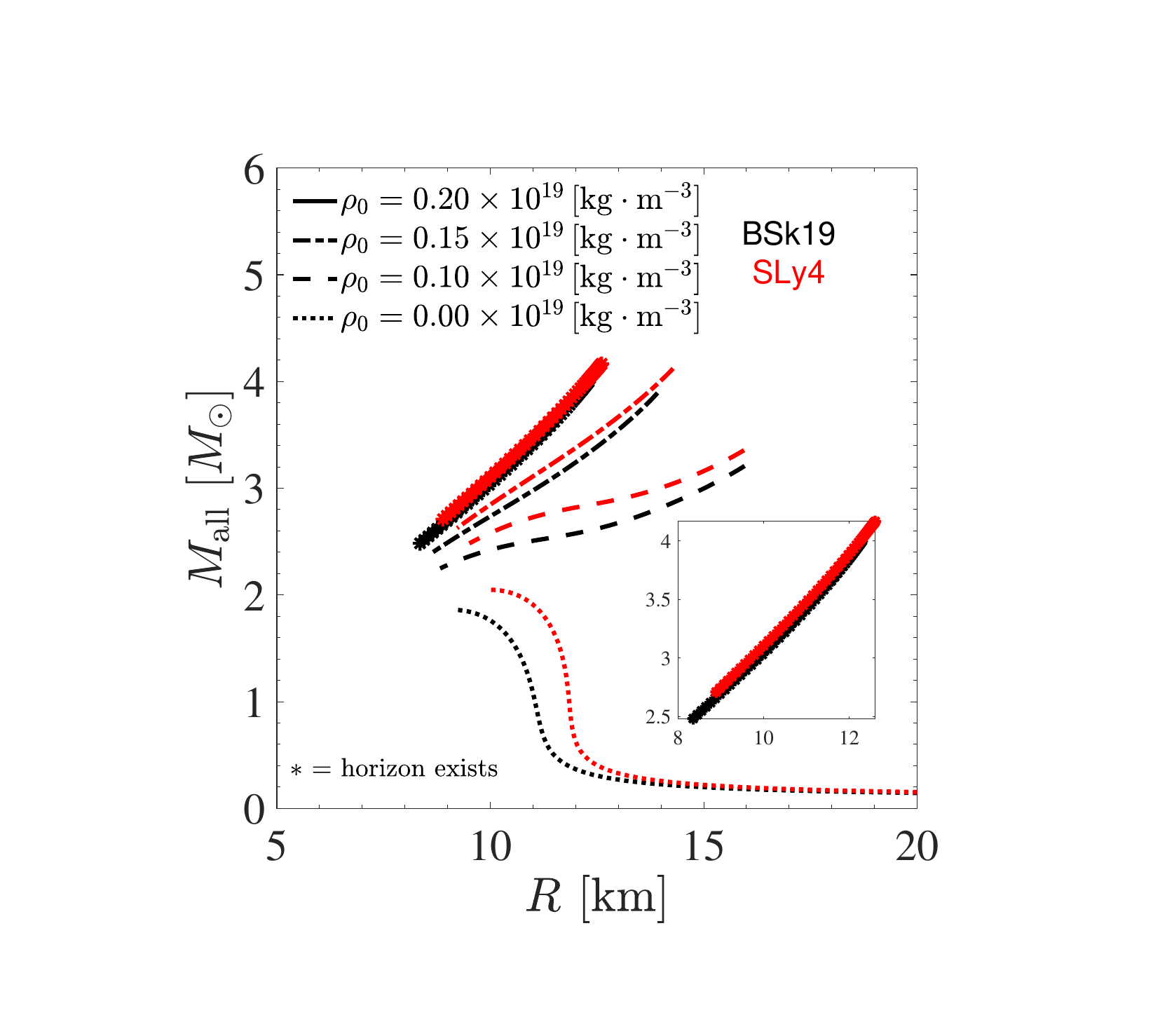}\label{fig:mr_total_n15}}
\hfill
\subfloat[Neutron star mass $M_{\rm NS}$ alone.]{\includegraphics[width=0.48\textwidth]{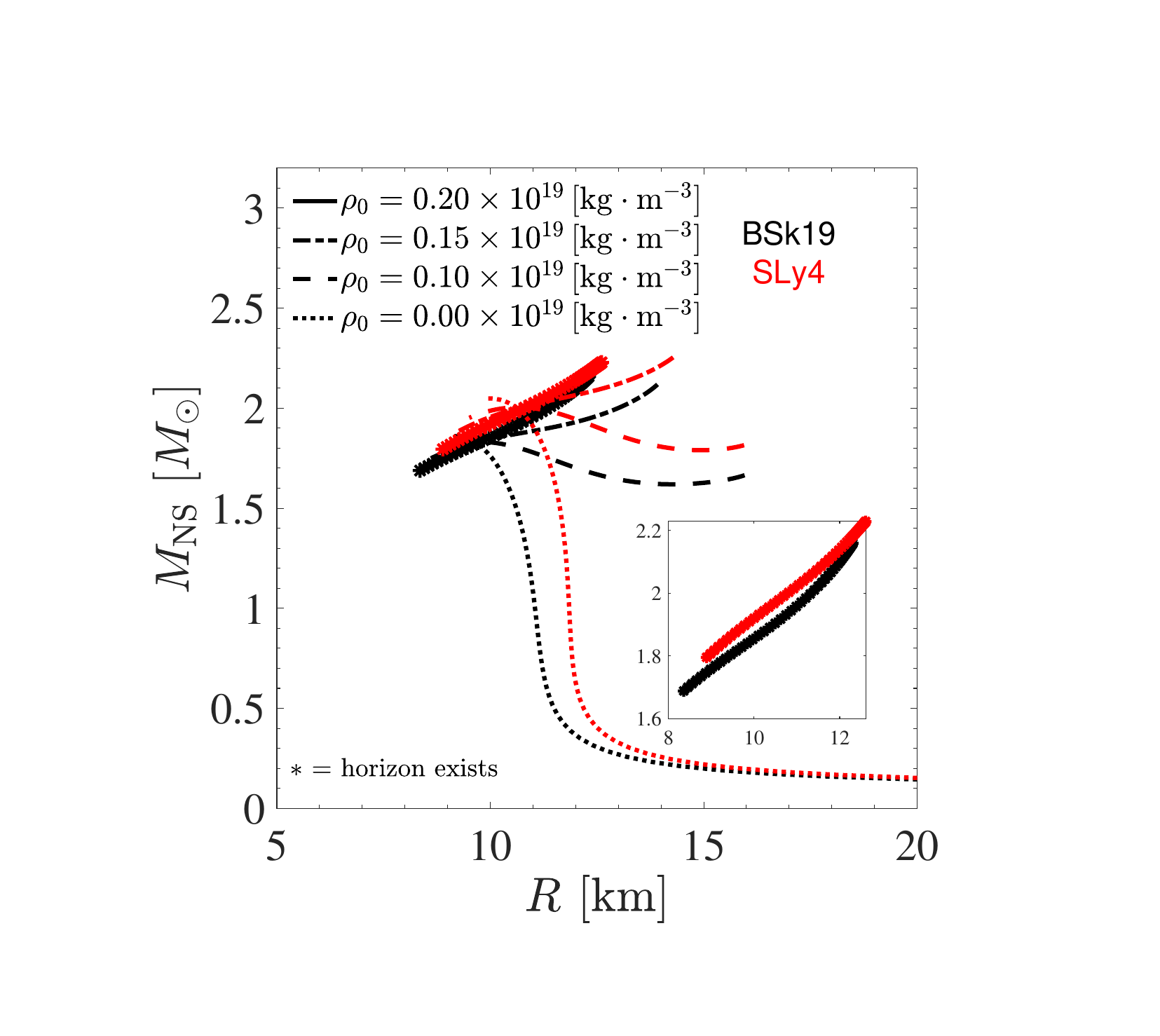}\label{fig:mr_ns_n15}}
\caption{Mass--radius relations for $n=1.5$ and $h=5~[\rm{km}]$. Solid (BSk19) and dashed (SLy4) curves correspond to different dark matter central densities $\rho_0$. Asterisks mark configurations for which $\min(g_{rr}^{-1})<0$, indicating the existence of a horizon.\label{fig:mr_combined_n15}}
\end{figure*}

\begin{figure}[t]
\centering
\includegraphics[width=\columnwidth]{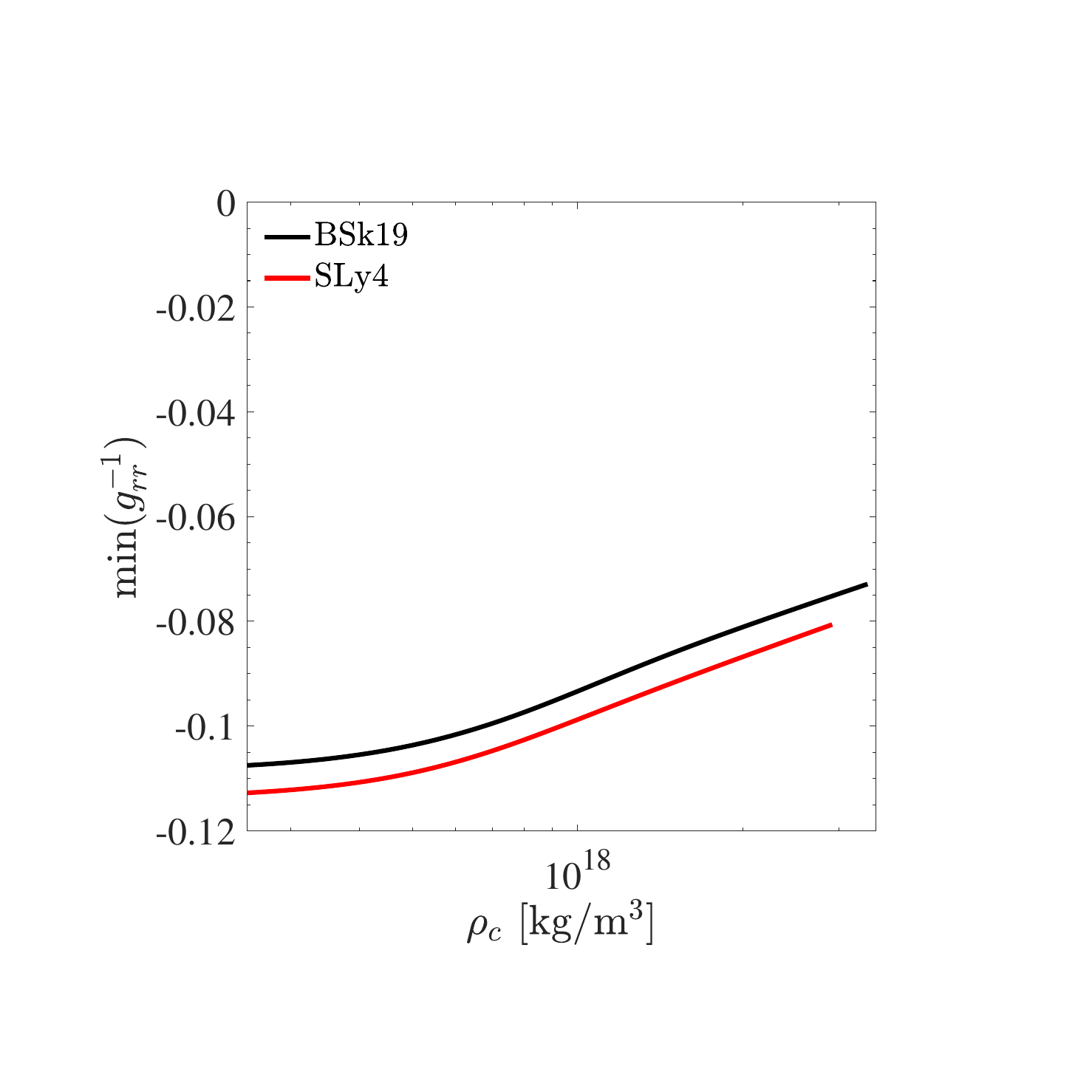}
\caption{Global minimum of $g_{rr}^{-1}$ as a function of the central baryonic matter density $\rho_c$, for $\rho_0=0.20\times10^{19}\,[\rm{kg/m^3}]$, $h=5~[\rm{km}]$, $n=1.5$.\label{fig:fmin_ed_n15}}
\end{figure}

For $n=1.5$, the shallower Einasto profile produces a stronger effect, as shown in Fig.~\ref{fig:mr_combined_n15}. This is further illustrated in Fig.~\ref{fig:fmin_ed_n15}, where, with $\rho_0=0.20\times10^{19}~[\rm{kg/m}^3]$, every configuration on the M-R curve is a neutron star in a black hole, i.e., $\min(g_{rr}^{-1})<0$ across the entire range of central densities for which static solutions exist.

The corresponding M-R relations for $n=1.5$ with $h=5~[\rm{km}]$ are shown in Fig.~\ref{fig:mr_combined_n15}, and representative values are listed in Tab.~\ref{tab:mr_summary_n15}. Owing to the shallower Einasto profile, a horizon already appears at $\rho_0=0.20\times10^{19}~[\rm{kg/m}^3]$, a considerably lower dark matter density than the $n=1$ case. The broader trend---the increase of $M_{\rm all}$ with $\rho_0$ and the consistency between BSk19 and SLy4---persists, confirming that the three-regime structure and the role of the dark matter halo in horizon formation are robust features of the model.

Tab.~\ref{tab:mr_summary} and Tab.~\ref{tab:mr_summary_n15} collect representative values at the maximum-mass point on each M-R curve for $n=1$ and $n=1.5$, respectively. At the horizon-forming densities ($\rho_0=0.40\times10^{19}~[\rm{kg/m}^3]$ for $n=1$ and $\rho_0=0.20\times10^{19}~[\rm{kg/m}^3]$ for $n=1.5$), $M_{\rm DM}$ is comparable to $M_{\rm NS}$, indicating that this configuration arises from the combined gravitational contribution of both components.
\begin{table}[t]
\caption{Representative values at the maximum of $M_{\rm all}$ on each M-R curve, for BSk19 with $h=5~[\rm{km}]$ and $n=1$.\label{tab:mr_summary}}
\begin{tabular}{c c c c c}
\hline\hline
$\rho_0$ [$10^{19}\,\rm{kg/m}^3$] & $M_{\rm all}$ [$M_\odot$] & $M_{\rm NS}$ [$M_\odot$] & $M_{\rm DM}$ [$M_\odot$] & $R$ [km] \\
\hline
0.00   & 1.8605 & 1.8605 & 0.0000 & 9.1553 \\
0.10  & 2.2286 & 1.7405 & 0.4881 & 9.7344 \\
0.20  & 2.8997 & 1.5709 & 1.3287 & 11.8133 \\
0.40  & 2.9319 & 1.3445 & 1.5874 & 8.6587 \\
\hline
\end{tabular}
\end{table}

\begin{table}[t]
\caption{Representative values at the maximum of $M_{\rm all}$ on each M-R curve, for BSk19 with $h=5~[\rm{km}]$ and $n=1.5$. A horizon first appears at $\rho_0=0.20\times10^{19}~[\rm{kg/m}^3]$.\label{tab:mr_summary_n15}}
\begin{tabular}{c c c c c}
\hline\hline
$\rho_0$ [$10^{19}\,\rm{kg/m}^3$] & $M_{\rm all}$ [$M_\odot$] & $M_{\rm NS}$ [$M_\odot$] & $M_{\rm DM}$ [$M_\odot$] & $R$ [km] \\
\hline
0.00 & 1.8605 & 1.8605 & 0.0000 & 9.1553 \\
0.10 & 3.2090 & 1.6649 & 1.5440 & 15.9583 \\
0.15 & 3.8929 & 2.1187 & 1.7742 & 13.9177 \\
0.20 & 4.0021 & 2.1593 & 1.8428 & 12.3052 \\
\hline
\end{tabular}
\end{table}

\subsection{Numerical Robustness}

The configuration of neutron stars in black holes is sufficiently exotic that one may ask whether the negative sign of $g_{rr}^{-1}$ could be an artifact of the numerical computation. We have verified its robustness by varying the ODE solver parameters over several orders of magnitude.

\begin{enumerate}
\item \textbf{ODE solver tolerance.} Varying $\text{RelTol}$ from $10^{-4}$ to $10^{-12}$---a factor of $10^8$, with $\text{AbsTol} = \text{RelTol} \times 10^{-4}$---changes $g_{rr,{\rm min}}^{-1}$ by at most $\sim 10^{-11}$, while the absolute changes in $R$ and $M_{\rm NS}$ are at most $\sim 10^{-6}~[\rm{km}]$ and $\sim 10^{-10}~[M_\odot]$.
\item \textbf{Maximum integration step.} Varying $\Delta r_{\rm max}$ from $0.1$ to $100~[\rm{m}]$---a factor of $10^3$---changes $g_{rr,{\rm min}}^{-1}$ by at most $\sim 10^{-14}$, with absolute changes in $R$ and $M_{\rm NS}$ at most $\sim 10^{-11}~[\rm{km}]$ and $\sim 10^{-13}~[M_\odot]$.
\item \textbf{ODE solver.} Replacing the adaptive fourth-order Runge--Kutta method with an implicit adaptive-step stiff solver (variable-order, 1--5) at $\text{RelTol}=10^{-8}$ changes $g_{rr,{\rm min}}^{-1}$ by $\sim 10^{-11}$. The absolute changes in $R$ and $M_{\rm NS}$ are at most $\sim 10^{-5}~[\rm{km}]$ and $\sim 10^{-10}~[M_\odot]$.
\end{enumerate}

Given the numerical stability of the solution and the fact that this configuration occupies a relatively broad parameter space (Fig.~\ref{fig:phase_diagrams}), these together indicate that the configuration of neutron stars in black holes is not an artifact of the numerical computation.

\section{Summary and Discussion}
\label{sec:discussion}

In this work, we have investigated the structure of neutron stars embedded in dark matter halos described by the Einasto density profile with the anisotropic equation of state $p_r = -\rho c^2$.
This dark matter model was recently shown by Konoplya and Zhidenko~\cite{Konoplya:2025ect} to produce regular, singularity-free black hole solutions.
By introducing a realistic neutron star equation of state (BSk19 and SLy4) that coexists with the dark matter halo and solving the modified Tolman--Oppenheimer--Volkoff equations, we have shown not only that neutron star solutions persist, but also that the dark matter halo profoundly alters the neutron star structure, and we have discovered a novel configuration: ``Neutron Stars in Black Holes.''

We first examined the non-monotonic influence of the three Einasto density profile parameters on the structure of dark matter admixed neutron stars.
We then demonstrated how the configuration of neutron stars in black holes gradually emerges, and mapped its parameter space.
We further computed the mass--radius relations for two different equations of state, finding that they deviate significantly from the predictions for ordinary neutron stars.
We also found that this configuration does not depend on a specific choice of the equation of state, and that the two components contribute comparably to its formation.
We further analyzed the numerical robustness of the solution, demonstrating that the result is stable under variations of the ODE tolerance, the maximum integration step, and the choice of solver.
Given the numerical stability and the fact that this configuration occupies a relatively broad parameter space (Fig.~\ref{fig:phase_diagrams}), these configurations of neutron stars in black holes are genuine solutions, not numerical artifacts.

More broadly, this work first clarifies how this distinctive dark matter, which can serve as a source for regular black holes, affects the structure of dark matter admixed neutron stars.
More importantly, the discovery of the configuration of neutron stars in black holes provides a new perspective: the interior of an event horizon need not be the end of physics.
Regular black holes can contain baryonic matter and support structure.
This offers a concrete, computable instance for fundamental questions about what lies inside a black hole and whether structure behind the horizon can leave observable imprints on external signals, providing a foundation for further investigation.

We also recognize that the dark matter densities required for the formation of the configuration of neutron stars in black holes far exceed the densities generally predicted for dark matter in the existing literature. Our study is therefore primarily a theoretical discussion, offering a fresh perspective and a concrete example for relevant research.
The dark matter model proposed in Ref.~\cite{Konoplya:2025ect} as a source for constructing regular black holes itself deviates from realistic astrophysical environments, and whether this model can be generalized to scenarios more consistent with current predictions for dark matter remains to be further explored. In the current model, the equation of state $p_{r{\rm (d)}} = -\rho_{\rm d} c^2$ requires a sufficiently large density to produce the negative pressure and mass effects needed to alter the neutron star structure; it has been shown that the equation of state for similar dark matter halo constructions can be further generalized ~\cite{Yue:2026evf}, and considering other equations of state may relax the demand on the dark matter density.

And several key questions merit further investigation.
The stability of these configurations under radial perturbations remains to be analyzed.
Most importantly, whether the configuration of neutron stars in black holes can be distinguished from a conventional Schwarzschild black hole, and whether effective observational signatures exist, remains an open question that will require substantial work to address.

\begin{acknowledgments}
This work is supported by the National Key Research and Development Program of China (Grant No. 2022YFC2204101 and 2020YFC2201503) and the National Natural Science Foundation of China (Grant No. 12275110 and No. 12247101).
\end{acknowledgments}

\end{document}